\documentclass[letterpaper, 10pt]{article}
\usepackage[margin=2.5cm]{geometry}

\usepackage[pdftex]{graphicx}

\usepackage[english]{babel}
\usepackage[T1]{fontenc}
\usepackage[utf8]{inputenc}

\usepackage{amsmath}
\usepackage{textcomp}
\usepackage{gensymb}

\usepackage{color,soul}
\usepackage{multirow}
\usepackage[noadjust]{cite}
\usepackage{subfig}
\usepackage[all]{nowidow}

\usepackage[pdfencoding=auto, pdfusetitle,hidelinks]{hyperref}

\begin{document}

\hypersetup{
  pdftitle={Smartphone-based user positioning in a multiple-user context with Wi-Fi and Bluetooth},
  pdfauthor={Viet-Cuong Ta, Trung-Kien Dao, Dominique Vaufreydaz, Eric Castelli},
  pdfkeywords={indoor localization, indoor navigation, collaborative positioning, Wi-Fi, Bluetooth, smartphone applications},
}

\title{Smartphone-based user positioning in a multiple-user context with Wi-Fi and Bluetooth}

\author{Viet-Cuong Ta$^1$, Trung-Kien Dao$^2$, Dominique Vaufreydaz$^3$, Eric Castelli$^3$}

\date{
{\small{\textsuperscript{1}~Human Machine Interaction, University of Engineering and Technology, Vietnam National University, Hanoi\\
\textsuperscript{2}~MICA Institute (HUST-Grenoble INP), Hanoi University of Science and Technology, Vietnam\\
\textsuperscript{3}~Univ. Grenoble Alpes, CNRS, Inria, Grenoble INP, LIG, 38000 Grenoble, France}}\\
\vspace{0.5em}\textit{Author version}
}

\maketitle

\begin{abstract}
In a multi-user context, the Bluetooth data from the smartphone could give an approximation of the distance between users. Meanwhile, the Wi-Fi data can be used to calculate the user's position directly. However, both the Wi-Fi-based position outputs and Bluetooth-based distances are affected by some degree of noise. In our work, we propose several approaches to combine the two types of outputs for improving the tracking accuracy in the context of collaborative positioning. The two proposed approaches attempt to build a model for measuring the errors of the Bluetooth output and Wi-Fi output. In a non-temporal approach, the model establishes the relationship in a specific interval of the Bluetooth output and Wi-Fi output. In a temporal approach, the error measurement model is expanded to include the time component between users' movement. To evaluate the performance of the two approaches, we collected the data from several multi-user scenarios in indoor environment. The results show that the proposed approaches could reach a distance error around 3.0m for 75 percent of time, which outperforms the positioning results of the standard Wi-Fi fingerprinting model.
\end{abstract}

\vspace{2em}
\begin{changemargin}{0.9cm}{0.9cm} 
\textbf{Keywords:} indoor localization, indoor navigation, collaborative positioning, Wi-Fi, Bluetooth, smartphone applications.
\end{changemargin}
\vspace{2em}

\section{Introduction}

In a GPS denied environment, Wi-Fi and Bluetooth could be considered as alternative wireless-based solutions for positioning purpose. Novel Wi-Fi-based positioning methods on smartphones can find the position by scanning the available Wi-Fi access points in the surrounding environment. The mean distance error is around 5m \cite{torres2017smartphone}, due to the unreliable characteristics of the Wi-Fi signal propagation in indoor environment.
In the case of Bluetooth-based positioning, the Bluetooth technology available on smartphones nowadays is much similar to the Wi-Fi technology in terms of underlying radio physical characteristics and application level. Therefore, it is possible to create a positioning system similar to the Wi-Fi ones. However, the Bluetooth communication range is smaller than that of Wi-Fi. To be of interest in a large area, it requires to deploy a high number of static beacons \cite{faragher2014analysis}. 

When several users are present, the Bluetooth data can be joined with the Wi-Fi data to create a collaborative positioning framework. When each user moves with {a} smartphone in a public area, it is possible to keep the smartphone’s Bluetooth in visible mode. Then, if a device sees another device nearby, the Receive Signal Strength (RSS) from Bluetooth data could give an approximation of the relative range between the two devices. Given the estimating pair-to-pair distance, it is possible to refine the output positions from Wi-Fi data. This approach does not require to install additional infrastructures and is compatible with the standard Bluetooth protocol which is generally supported by smartphones.

There are several key challenges of the proposed approach. The first difficulty is the noisy propagation characteristics of radio signals in indoor environment. The noise affects both Wi-Fi output positions and Bluetooth output distances, moreover in the case of moving users. The second difficulty is the mis-synchronization between the Wi-Fi and Bluetooth scanning processes in the smartphone. In other words, the scanning cycle for each technology in the smartphone are not guaranteed to start and finish at the same time.

In this work, we try to overcome these problems by considering {non-temporal and temporal approaches}. In the non-temporal approach, the distance error of the Wi-Fi output position is modeled by a Gaussian distribution. Similarly, another Gaussian distribution is used to describe the distance error between two devices from the Bluetooth inquiry process. Wi-Fi and Bluetooth outputs within a short time period are treated as if they happen in a same time window. An error function is then created to measure the mismatch between the two distributions. By neutralizing the mismatch, it is possible to improve the position results of Wi-Fi output. In the temporal approach, the time component of the users’ movement is incorporated into the error function. The error function uses the positions of all the users as parameters. We employ particle-filter-based tracking to minimize the error function. The particle filter uses as observation model the combination between Wi-Fi and Bluetooth scanning data. The experiments were conducted with real scenarios with up to four users. Both the non-temporal and temporal approach results are tested against the standard Wi-Fi fingerprinting model. Our results show that it is possible to make use of Bluetooth signal to improve the positioning output of the Wi-Fi fingerprinting model.

The remaining parts of the paper are arranged as follow: in the section II, the related works on smartphone-based indoor positioning and collaborative positioning are introduced. Our approaches for combining Wi-Fi and Bluetooth data are presented in section IV. The experiments and results are carried out in section IV. The section V contains the conclusion and future works.

\section{Related works}

The usage of Wi-Fi for positioning is well-studied. Popular approaches included geometry-based approaches \cite{bose2007practical,chintalapudi2010indoor} or fingerprinting based approaches \cite{he2016wi}. The fingerprinting based are preferred because it takes benefit from the deployment of WLAN infrastructure. Recently works on smartphone-based with Wi-Fi fingerprinting have reached a mean distance error of around 5m \cite{torres2017smartphone,mathisen2016comparative}. For learning techniques, the well-known K-Nearest Neighbors (KNN) and its alternative are among the most popular technique \cite{ma2015improved,liu2016smallest}. In \cite{torres2016ensembles}, the authors differ a wide range of KNN parameters to get a set of models. An ensemble result of the generated KNN models has a mean distance error of around 6m. Besides KNN-based learning methods, decision tree-based learning methods can be used for learning the Wi-Fi signal characteristics with prominent results \cite{ta2016smartphone}.

For positioning purpose, Bluetooth technology can be employed in the same way with Wi-Fi technology. Bandara et al. \cite{bandara2004design} use up to four Bluetooth antennas as static stations. The proposed system is able to locate a Bluetooth tag within a room with area of 4.5m \( \times \) 5.5m. The RSSI value is used to classify the tag’s position between different subareas of the room. Pei et al. \cite{pei2010using} employ fingerprinting-based approach to track a moving phone. The setup includes only three Bluetooth beacons in a corridor-like space of 80m long approximately. The horizontal error is reported at 5.1m. For comparison, the Wi-Fi-based solution has an error of 2.2m in the same area. However, these results are possible thanks to the 8 installed WLAN access points. More recent works employ the new BLE technology. The BLE beacons are smaller and more energy efficient. They are able to power up for a longer period of time \cite{gomez2012overview}. Thus, they are more convenient to create Bluetooth beacon networks for positioning purpose. Faragher and Harle \cite{faragher2014analysis} provide an in-depth study of using BLE for indoor localization purpose. The distance error of Bluetooth-based approach could reach as low as 2.6m for 95\% of times. However, a high number of beacons should be deployed to reach the above performance. The study also addresses some issues of the BLE signal such as the scanning cycle, fast fading effects and Wi-Fi scanning interference. A similar performance for BLE-based indoor positioning is reported in \cite{zhuang2016smartphone}. The authors employed fingerprinting-based approach with the RSS value from the installed BLE beacons. 

In a scenario involving multiple devices, there are several works on collaborative localization. Those works rely on some specific wireless technologies, which support the peer-to-peer communication. These technologies include Bluetooth, Wi-Fi Direct and Sound. They are capable to discover the existence of nearby neighbors. In \cite{liu2014face}, the task of detecting face-to-face proximity is studied. The smartphones are used to scan nearby visible Bluetooth devices in daily usage. From the received RSSI, relative distance between two devices is calculated. The distance is then used to detect whether the two users are closed to each other. To deal with noisy Bluetooth signals, additional techniques such as RSSI smoothing and light sensor data are introduced for calculating a more accurate distance. \cite{jun2013social} propose the Social-Loc system, which uses Wi-Fi Direct technology for detecting two events: \textit{Encounter} and \textit{Non-Encounter} between each pair of users. In their work, the authors find the RSSI peak for separating Encounter and Non-Encounter events. These detected events are then used to improve the Wi-Fi fingerprinting and Dead Reckoning tracking. The drawback of Wi-Fi Direct technology is that it does not allow a regular Wi-Fi scanning. Therefore, the proposed Social-Loc is more suitable for improving the Dead Reckoning tracking than the Wi-Fi fingerprinting tracking. Sound-based ranging is also useful to detect the relative distance between two devices. In \cite{liu2012push}, the authors use the sound-based distance to improve Wi-Fi fingerprinting-based positioning system. The acoustic ranging is designed with TOA method for calculating the distance between devices. The estimated ranges are then used to form a graph between devices. The graph’s vertices are derived from the Wi-Fi positioning output. A search within the graph is then performed to find the best match position. The searching task aims to find an agreement between the vertices’ position and the edges’ length. The proposed approach has a mean error of around 1.6m, depending on specific setups. However, the study only mentions the cases when all the devices are in static position.

\section{Using Bluetooth Data to Improve Wi-Fi Positioning}

Wi-Fi data and Bluetooth data are two data streams which carry different information of users' position. For indoor positioning, the positioning of a user can be derived the Wi-Fi technology by scanning RSS signal of nearby access points. When there are multiple users, the distance from one user to other user can be calculated from the Bluetooth scanning process. A way to combine the two different data streams is to use an additional central server. The server keeps all the available positioning information from each participate devices. More precisely, the information includes the estimated position from Wi-Fi data and the estimated distances between pairs of devices from Bluetooth-data. There are several works in the literature which take the same approach for indoor positioning based on fusing different data streams. For example, \cite{dao2014user} uses a server-based solution for combining different information to improve the localization results. 

\subsection{Centralized Positioning Framework with Wi-Fi and Bluetooth}
In our task of fusing Wi-Fi and Bluetooth data, two types of required information must be send to the server side. The first type is the scanned Wi-Fi information of access points. For each completed scan cycle, the device sends identifier (Wi-Fi MAC address) of the seen access points and their RSS values. Each scan cycle lasts several seconds and is device dependent. The second type is the Bluetooth scanned information. The scanned information contains the Bluetooth MAC addresses of seen devices and their RSS values. The time of a complete Bluetooth scan is not defined. When a new Bluetooth device is seen, it could be sent to the server immediately. In practice, there can be many visible Bluetooth devices within the environment such as wireless headphones or mice. The server side maintains a list of active devices. From the list, only the Bluetooth information from the participant devices is kept for positioning purpose.

\begin{figure}
\centering
\includegraphics[scale=0.7]{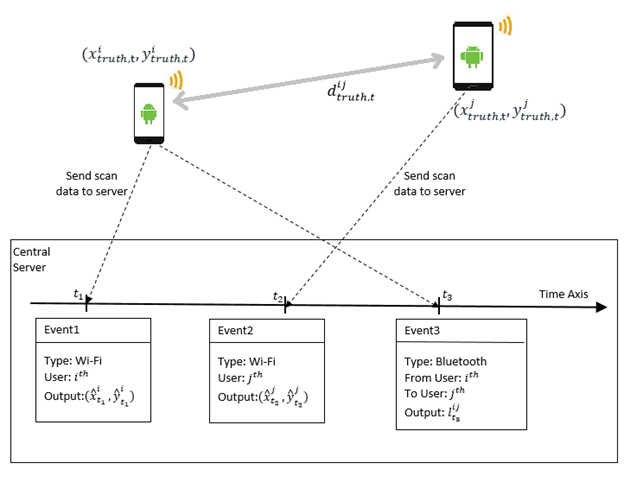}
\caption{The data send from two devices to the central server and the derived information at the server for estimating each devices' position}
\label{fig::systemoverview}
\end{figure}

On the server side, the data from Wi-Fi and Bluetooth scans give different ways to calculate the users’ positions. Figure \ref{fig::systemoverview} illustrates the principle of our approach. For simplicity, we consider the positioning problem within a single floor. Each user is identified by his smartphone. The example context involves two users, namely the  $i^{th}$ and $j^{th}$ users. The two users’ devices keep gathering Wi-Fi access points data and Bluetooth inquiry data within the environment and send {them} to the server. The server receives the data and stores {them as typed Events (Wi-Fi or Bluetooth)}. In Figure \ref{fig::systemoverview}, there are three Events: two Wi-Fi scans and a Bluetooth scan. The real positions of users $i^{th}$ and $j^{th}$ are denoted as $(x_{truth,t}^i,y_{truth,t}^i)$ and $(x_{truth,t}^j,y_{truth,t}^j )$, respectively. The subscript $t$ is the timestamp. The real distance between the two users is  $d_{truth,t}^{ij}$.

At time $t_1$, {one} can determine the position of user $i^{th}$ as $(\hat{x}_{t_1}^i,\hat{y}_{t_1}^i)$ from the WLAN scan information of the $i^{th}$ device. The Wi-Fi position output, however, could be different from the real position $(x_{truth,t_1}^i,y_{truth,t_1}^i)$ of the user. Similarly, at time $t_2$, when the $j^{th}$ device completes a Wi-Fi scan, we can compute the Wi-Fi position output of user $j^{th}$. Let the result of this computation be $(\hat{x}_{t_2}^i,\hat{y}_{t_2}^i)$. Besides that, the Bluetooth scanning process could give an estimated distance between the two users. At time $t_3$, if the two users $i^{th}$ and $j^{th}$ are within the Bluetooth scanning range, we could find the relative distance $l_{t_3}^{ij}$ from the RSS value of Bluetooth scanning process. The value of $l_{t_3}^{ij}$ is an approximation of the real distance $d_{truth,t_3}^{ij}$ between the two users at time $t_3$. An alternative way for calculating $d_{truth,t_3}^{ij}$ is using the output position from the Wi-Fi data for both users $i^{th}$ and $j^{th}$. 

In order to benefit from the relationship between Wi-Fi and Bluetooth for improving positioning results, we {question} two different approaches. The first one is a \textit{Non-temporal} approach. The temporal relationship between events which are within a time interval window is removed. They are treated as they happened at the same time. The Wi-Fi fingerprinting approach is used for finding user’s positioning from Wi-Fi scan. The Log Distant Path Loss (LDPL) model \cite{rappaport1996wireless} is used to find the distance from the input Bluetooth RSS value. An error estimation function is established by using the users’ position as the function's parameters. By minimizing the error function, ir can be used to smooth the mismatch between Wi-Fi data and Bluetooth data, thus, to reduce the positioning error from Wi-Fi. The second approach is the \textit{Temporal} one. We introduce the time component into the basic error function of the first approach. More precisely, the new error function includes the devices’ position at each timestamp. For minimizing the new error function, a particle-based approximation is carried out. 

\subsection{Non-temporal Approach}

In the \textit{Non-temporal} approach, a sliding window of length $\Delta t$ is used. All the events from $t$ to $t+\Delta t$ are considered to happen at the same time. Given a pair of user $i^{th}$ and $j^{th}$, {one} can assume that there exists both Wi-Fi data and Bluetooth data within the time frame from $t$ to $t + \Delta t$. Let $w^i$ and $w^j$ be the Wi-Fi scans from the two users, and $rss^{ij}$ is the RSS value of the Bluetooth scan. {In this \textit{Non-temporal} approach}, we remove the time variable $t$ from the parameters. {The} likelihood function with the two users’ position $(x^i,y^i)$ and $(x^j,y^j)$, and the parameters $P(x^i,y^i,x^j,y^j |w^i,w^j,rss^{ij})$, {is created} by splitting down the function into three separate components:
\begin{equation}
\begin{split}
P(x^i,y^i,x^j,y^j |w^i,w^j,rss^{ij}) = & P_W (x^i,y^i |w^i) \times \\ &
P_W (x^j,y^j |w^j) \times \\ & P_B (x^i,y^i,x^j,y^j |rss^{ij})
\label{eq::PNonTemporal}
\end{split}
\end{equation}
{In the equation}, $P_W (x^i,y^i |w^i), P_W (x^j,y^j |w^j)$ are the error estimation from Wi-Fi data of user $i^{th}$ and $j^{th}$ and $P_B (x^i,y^i,x^j,y^j |rss^{ij})$ is the error from the Bluetooth data.

Let $(\hat{x}^i,\hat{y}^i)$ be the {computed} position from the scan $w^i$. By assuming the distribution of the real position $(x^i,y^i)$ to be a 2D Gaussian around the estimated position $(\hat{x}^i,\hat{y}^i)$, $P_W (x^i,y^i |w^i)$, and is measured by:
\begin{equation}
P_W (x^i, y^i | w^i ) \sim \frac{1}{\sqrt{2 \pi} \delta_w}{e^{- \frac{(x^i- \hat{x}^i )^2+(y^i - \hat{y}^i )^2}{2\delta_w^2}}}
\label{eq::PWIFIi}
\end{equation}

Similarly, it is possible to compute $P_W (x^j,y^j |w^j)$, given the estimated position $(\hat{x}^j,\hat{y}^j)$ from Wi-Fi positioning models:
\begin{equation}
P_W (x^j, y^j | w^j ) \sim \frac{1}{\sqrt{2 \pi} \delta_w}{e^{- \frac{(x^j- \hat{x}^j )^2+(y^j - \hat{y}^j )^2}{2\delta_w^2}}}
\label{eq::PWIFIj}
\end{equation}

{To estimate} the Bluetooth part of the estimated likelihood, we first calculate $l^{ij}$ from $rss^{ij}$ by using the well-known LPDL model:
\begin{equation}
l^{ij} = l_0 \times 10^{\frac{rss^{ij}-rss_{l_0}}{10n}}
\label{eq::PBluetoothij}
\end{equation}
where $rss_{l_0}$ is the RSS value at the distance $l_0$, $n$ is the path loss exponent. The three values $rss_{l_0}$, $n$, and $l_0$ are known constants. The value of $l^{ij}$ is an approximation of the real distance, which comes directly from the real position of users $i^{th}$ and $j^{th}$,  $(x^i,y^i)$ and $(x^j,y^j)$:
\begin{equation}
d^{ij}=\sqrt{(x^i-x^j)^2-(y^i-y^j )^2}
\label{eq::distance}
\end{equation}

{One can assume} that $l^{ij}$ has a Gaussian distribution around $d^{ij}$, the Bluetooth likelihood can be estimated by another Gaussian kernel:
\begin{equation}
P_B (x^i,y^i,x^j,y^j |rss^{ij}) \sim \frac{1}{\sqrt{2 \pi} \delta_w}{e^{-\frac{(d^{ij}-l^{ij})^2}{2\delta_b^2 }}}
\label{equaion::bltrss}
\end{equation}
with $\delta_b$ is a constant indicating the reliability of the LDPL on the RSS Bluetooth signal.

From Equations \ref{eq::PWIFIi}, \ref{eq::PWIFIj} and \ref{eq::PBluetoothij}, the likelihood function in Equation \ref{eq::PNonTemporal} could be rewritten as:
\begin{equation}
P(x^i,y^i,x^j,y^j | w^i,w^j, rss^{ij} )= C \times e^{-g}
\end{equation}
with $C$ is an constant, $g$ is a function of $x^i,y^i,x^j,y^j$, and 
\begin{equation}
\begin{split}
g = & \frac{(x^i- \hat{x}^i )^2 + (y^i- \hat{y}^i )^2+(x^j-\hat{x}^j )^2+(y^j- \hat{y}^j )^2}{2\delta_w^2} + \\ &
\frac{( \sqrt{(x^i-x^j)^2-(y^i-y^j )^2 } -l^{ij} )^2} {2\delta_b^2 }
\end{split}
\end{equation}

For a fast computing of the minimum value of $g$, two constraints can be added {based on the symmetric properties} of $g$. The first constraint is that the four points $(x^i,y^i), (x^j,y^j), (\hat{x}^i, \hat{y}^i),(\hat{x}^j, \hat{y}^j)$ are aligned. The second constraint is that the distance between $(x^i,y^i)$ and $(\hat{x}^i, \hat{y}^i)$ and the distance between $(x^j,y^j)$ and ($\hat{x}^j, \hat{y}^j)$ are equal. The function is then rewritten as a function of the distance $r$ between $(x^i,y^i)$ and ($\hat{x}^i, \hat{y}^i)$, whose minimum value can be easily computed.
\begin{equation}
g(r)=\frac{r^2}{\delta_w^2}+\frac{(\hat{d}^{ij}-2r-l^{ij} )^2}{2\delta_b^2}
\end{equation}

The removing time information of the incoming Wi-Fi and Bluetooth data make the error probability can be approximated by the function $g$. However, it introduces several drawbacks. The first one is {due} to the users’ movement. In different time windows $\Delta t$, the Wi-Fi position and the Bluetooth-based distance{s are variable}. The second drawback is the difficulty to determine the time interval length $\Delta t$ for grouping consecutive Wi-Fi data and Bluetooth data. The later temporal approach is designed to overcome those drawbacks of the non-temporal approach.

\subsection{Temporal Approach}

In our \textit{Temporal} approach to the problem, we attempt to use the temporal relationship in the likelihood function $P$. Instead of relying only on the position of two users at specific timestamp to measure the errors, the likelihood function could be extended to include the moving path of all the users. Each moving path is considered as a sequence of points. The new likelihood function would receive all the points as parameters.

We first construct the likelihood function $F$, which is a more {complete} form of $P$ that bases on three probability functions. A motion model $M$ is used to establish the relationship between the position at time $t$ and the position at time $t+1$ when the user moves within the area. A probability distribution function $W$ describes the distribution probability from Wi-Fi scan results. The $B$ function describes the distance distribution based on the Bluetooth RSS value from each pair of devices.

Let assume that there are $N$ users to track in $T$ seconds. The time component is added to the position of user $i^{th}$ at time $t$ as $(x_t^i,y_t^i)$. Normally, we can select the time delta value based on the specific purpose of the positioning system. For modeling purpose, it is required that the time index $t$ contains all the event timestamps from Wi-Fi and Bluetooth of all the participant devices. Approximately, all the float-typed timestamps could be rounded to the nearest integer values. The motion model for each user $i^{th}$ is defined as a probability function between the previous position and {the} present position, $M(x_t,y_t | x_{t-1},y_{t-1})$. The moving component for $T$ seconds for each user $i^{th}$ is then calculated by:
\begin{equation}
F_i^M = \prod_{t=1}^T M(x_t^i,y_t^i | x_{t-1}^i, y_{t-1}^i)
\end{equation}

For each specific user $i^{th}$, assum{ing} that there are $K$ timestamps {within} the $T$ seconds which have the Wi-Fi scan results. One can set the timestamps for Wi-Fi events as $u_1$,$u_2$,...,$u_K$. Then for each $u^k$, the Wi-Fi data is denoted as $w_(u^k)^i$, the function $W$ is used to estimate the likelihood probability $W(x_{u_k}^i,y_{u_k}^i | w_{u_k}^i)$. Then, the Wi-Fi component $F_i^W$ of the user $i^{th}$ is built from all the available $K$ Wi-Fi scans:
\begin{equation}
F_i^W = \prod_{k=1}^K W(x_{u_k}^i,y_{u_k}^i | w_{u_k}^i)
\end{equation}

The Bluetooth evolves {for} a specific pair of user $i^{th}$ and user $j^{th}$. Assuming that there are $L$ Bluetooth data which arrive at timestamps $v_1$,$v_2$,...,$v_L$, it is possible to chain the error function over the $L$ timestamp as follow:
\begin{equation}
F_{ij}^B = \prod_{l=1}^L B(x_{v_l}^i,y_{v_l}^i,x_{v_l}^j,y_{v_l}^j | rss_{v_l}^{ij})
\end{equation}

By joining the three functions, the total likelihood function $F$ could be written as:
\begin{equation}
F=(\prod_{i=1}^N F_i^M)(\prod_{i=1}^N F_i^W)(\prod_{i=1,j=1}^N F_{ij}^B)
\end{equation}
At this step, one could select the explicit form of $M$, $W$ and $B$ and process them to find the maximum value of $F$. The number of estimated parameters in $F$ totally depends on the number of users and the tracking time. As the $F$ function includes a motion model $M$, particle filter-based approximation is a natural way for approximating the maximum value of $F$. In addition to that, the particle filter process would have a more flexible way for selecting the explicit forms of $W$ and $B$.

\begin{figure}
\centering
\includegraphics[scale=1.0]{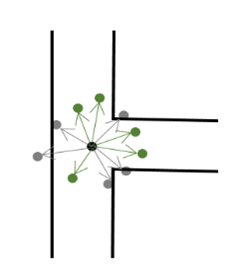}
\caption{A simple moving model with the black dot is the initial particle. \label{fig::motionmodel}}
\end{figure}

For each user $i^{th}$ at time $t$, there is a set of particles $S_i^t$, which represents the position distribution probability. For the motion model $M$, without the additional information from inertial sensors, the movement of the user could take random values as the moving speed $v$ and the heading direction $h$. While there is no constraint on the value of $h$, the moving speed $v$ should be suitable with a typical indoor movement. In our specific implementation, we generate the moving speed from a normal distribution around a speed average value. The speed average is chosen according to the walking action in indoor environment. The heading is generated from the uniform distribution in the range $[0,2\pi]$. An additional wall-crossing checking step is added for removing bad particles. Figure \ref{fig::motionmodel} gives an example of the motion model $M$ for generating the new particles. The center black dot is the initial particle. The walls are represented with black lines. New particles are then generated with a normal distribution moving speed around the speed average value and a uniform heading direction. The green particles are kept. The gray ones, which cross the wall, are removed.

With the particle filter-based approximation, the likelihoods given by Wi-Fi component $F^W$ and Bluetooth component $F^B$ could be transformed into an observation model. The score a specific $p_i^t \in S_i^t$ is calculated by:
\begin{equation}
score(p_i^t)=score_W (p_i^t)+score_B (p_i^t)
\end{equation}
with $score_W (p_i^t)$ is the Wi-Fi component and $score_B (p_i^t)$ is the Bluetooth component.

If there exists a Wi-Fi scan $w$ at time t, the $score_W (p_i^t)$ is {then} computed by using a local estimation on the probability output of Wi-Fi fingerprinting model. The area is first divided into separated clusters, $C_1$, $C_2$, ...,$C_D$. Using the clusters, we can transform the Wi-Fi fingerprinting model from a standard regression problem into a classification problem \cite{ta2016smartphone}. Let $prob_w = \{a_1, a_2, .., a_D\}$ is the chance of the predicted output of $w$ belong to the clusters. $score_W (p_i^t)$ can be computed as followed:
\begin{equation}
score_W (p_i^t) = \sum_{i=1}^{D} score_{C_i}(p)
\end{equation}
where $score_{C_i}(p)$ is a scaled-value from $prob_w$, with respect from the maximum distance $dmax_{C_i}$ and minimum distance $dmin_{C_i}$ to all the available particles:
\begin{equation}
score_{C_i}(p) = a_i \times (1 - \frac{d(p, C_i) - dmin_{C_i}}{dmax_{C_i} - dmin_{C_i}})
\end{equation}
A constant $\Delta_1$ t is the effective window length for each Wi-Fi scan. 

Similarly, the $score_B (p_i^t)$ can be calculated if there is any Bluetooth scan involving the user $i^{th}$ around time $t$. Without loss of generality, we assume {that} the available Bluetooth scan is $rss_{ij}^t$ that specifies the RSS value from the device $i^{th}$ and $j^{th}$. The update rule for $score_B (p_i^t)$ is defined as follow:
\begin{equation}
score_B (p_i^t)=\sum_{p_{j,k}^t \in S_j^t}  score_W (p_{j,k}^t) \times B(p_i^t,p_{j,k}^t | rss_{ij}^t)  
\end{equation}
The subscript $k$ indicates the need to calculate repeatedly for each $p_{j,k} \in S_j^t$. The likelihood $B(p_i^t,p_{j,k}^t | rss_{ij}^t)$ is computed by using similar process as the computation of the likelihood $P_B$ in the \textit{Non-temporal} approach. Using the Equation \ref{equaion::bltrss}, the likelihood $B$ can be rewritten as a function of the distance $d$ is the distance between two particles $p_i^t,p_{j,k}^t$ and the distance $l$ is derived from $rss_{ij}^t$  by the LDPL model. A constant $\Delta_2 t$ is added to define the effective interval length for a Bluetooth scan.

\section{Experiment and Results}

\subsection{Experiment Setup}
\begin{figure}
\centering
\includegraphics[scale=0.75]{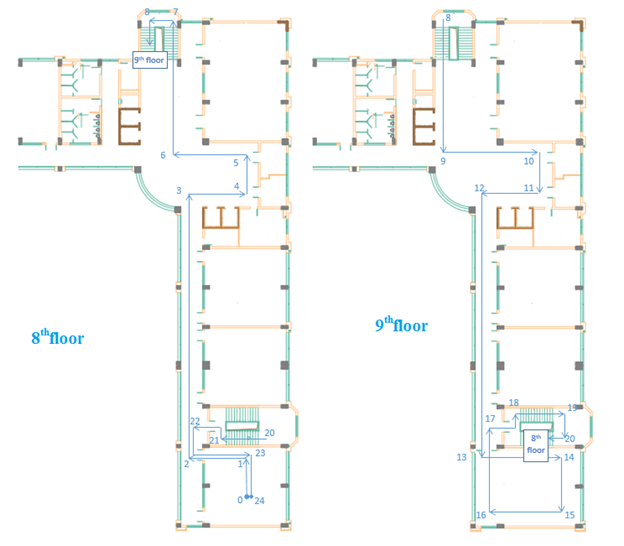}
\caption{The moving path, colored in blue color, which includes two floors\label{fig::movingpath}. The checkpoints are numbered along the moving path.}
\end{figure}

{To evaluate the performance of the two proposed approaches, we setup an experiment within an office environment which includes two floors.
All recordings follow one common trajectory which is composed by the corridor, office rooms and stairs.
This trajectory is defined by several checkpoints}.
{The path is illustrated in Fig.} \ref{fig::movingpath}.
{The length of the path is approximately 200m, which usually takes about 300s of walking at an average speed.}

{Different scenarios were designed in such a way that the additional Bluetooth scanning data could provide useful information for smoothing the Wi-Fi positioning output.
Each of them involves groups of 2 to 4 users.
The users were instructed to carry the devices and move around the experimental area.
They walk along the same path, with different relative distances between each other.
In the recordings, we used an interval of 0.5s for tracking the users.
Each time a checkpoint is reached, the time is registered.
The checkpoint's position and its reaching time are then used to calculate the user's trajectory as the ground truth movement.

We selected four devices including two smartphones and two tablets for the positioning scenario (see table} \ref{tab::testdevices}).
{Each device is set to scan Wi-Fi access points and available Bluetooth devices in the environment. All of them run Android operating system and use the same application for collecting the Wi-Fi and Bluetooth data.}

\begin{table}
\caption{The four participated devices in the testing scenarios \label{tab::testdevices} }
\centering
\begin{tabular}{|c|c|c|}
\hline
\textbf{Device ID} & \textbf{Name} & \textbf{Type} \\
\hline
1 & Samsung Galaxy Note 4 & Smartphone \\
\hline
2 & HTC One ME & Smartphone \\
\hline
3 & Asus ME & Tablet \\
\hline
4 & Samsung Galaxy Tab & Tablet \\
\hline
\end{tabular}
\end{table}

{Three approaches are used to localize the users when they are moving within the area: \textit{Wi-Fi only} (for comparison), \textit{Non-temporal} and \textit{Temporal}.}
In the \textit{Wi-Fi only} approach, the output of Wi-Fi fingerprinting method is provided as the reference tracking results. First, the data is collected over the walking path.
A Random Forest (RF) regressor model is trained on the collected data, using the method described in \cite{ta2016smartphone}. In the testing phase, with each completed Wi-Fi scan from the tested devices, the RF model is used to produce the position output. 

In the \textit{Non-temporal} approach, firstly, the output positions from the pre-trained RF regressor model on Wi-Fi are calculated for each device. If there is any Bluetooth data available, the Bluetooth scanned information is used for adjusting the positions of two involved devices. To solve the problem of non-simultaneous events between Wi-Fi scans and Bluetooth scans, we use a time window of length $\Delta t=10$ seconds for grouping successive events into the same timestamp.
The resulting position is calculated as the mean value of these positions.

In the \textit{Temporal} approach, a RF classifier model is built on top of the RF regressor model. To transform the real world coordinates to label index, we perform a K-means clustering of all the available training positions from the training data. The new learning targets are the indices of the corresponding clusters. In our experiments, we use $K = 30$ for clustering all the available points in the tested area. The radius of each cluster in this configuration is 4.0m approximately. The probability output of the classifier model is then used to update the Particle Filter within a time window of 10 seconds. If there are multiple completed scans within this time window, the nearest completed scan is selected. The Bluetooth data has the effective range set to 2.0 seconds. For the moving model, the average speed of each particle is set to 1m/s. In the simulation step, the number of particles is set to 1000.

\subsection{Results and Discussion}

The tracking results of three discussion approaches are illustrated in the Table \ref{table::results}. In the ``one group'' setups, every user is instructed to move as a group throughout the corridors. In other configuration, users are instructed to move with a distance under 10m to each other. The special case {is with 4 users: there are 2 groups of 2 users, which let the system use Bluetooth data to identify both closed and distant devices.} The results are reported as the mean average distance errors across all the testing devices in the specific scenario.
\begin{table}
\caption{Positioning errors as averages among different devices in several contexts, where the users are asked to move in different groups \label{table::results}}
\centering
\begin{tabular}{| c | c | c | c | c |}
\hline
\textbf{N of}  & \textbf{N of} & \multirow{2}{*}{\textbf{Wi-Fi Only}} & \multirow{2}{*}{\textbf{Non-temporal}} & \multirow{2}{*}{\textbf{Temporal}} \\
\textbf{Users} & \textbf{Groups} &  &  &  \\
\hline
2 & 1 & 3.7m $\pm$ 2.0m & 2.4m $\pm$ 1.6m & 2.2m $\pm$ 1.5m \\
\hline
2 & 2 & 3.4m $\pm$ 2.4m & 3.2m $\pm$ 2.3m & 2.5m $\pm$ 2.0m \\
\hline
3 & 1 & 4.0m $\pm$ 2.4m & 3.5m $\pm$ 2.2m & 2.1m $\pm$ 1.8m \\
\hline
3 & 3 & 3.6m $\pm$ 2.1m & 3.1m $\pm$ 2.2m & 2.2m $\pm$ 1.9m \\
\hline
4 & 1 & 3.8m $\pm$ 2.6m & 3.6m $\pm$ 2.6m & 2.8m $\pm$ 2.0m \\
\hline
4 & 2 & 3.8m $\pm$ 2.4m & 3.5m $\pm$ 2.3m & 2.6m $\pm$ 2.1m \\
\hline
\end{tabular}
\end{table}

The \textit{Wi-Fi Only} approach reaches a stable performance of under 4.0m in mean distance error. Both \textit{Non-temporal} and \textit{Temporal} approaches have better results than the \textit{Wi-Fi Only} approach. However, the \textit{Non-temporal} approach's results are not as stable as the \textit{Temporal} one. In the setup where the users' distance could change within a specific time interval, the \textit{Non-temporal} approach has similar performance as the output from the Wi-Fi fingerprinting model. In this case, it raises the difficulty to measure the distance in the Equation \ref{eq::distance}. Meanwhile, the \textit{Temporal} give a more stable performance. It can decrease the errors from 25\% to 50\% based on specific testing setups. The biggest relative improvement is the setup of three users moving in one group.

\begin{figure}
\centering
\includegraphics[scale=0.6]{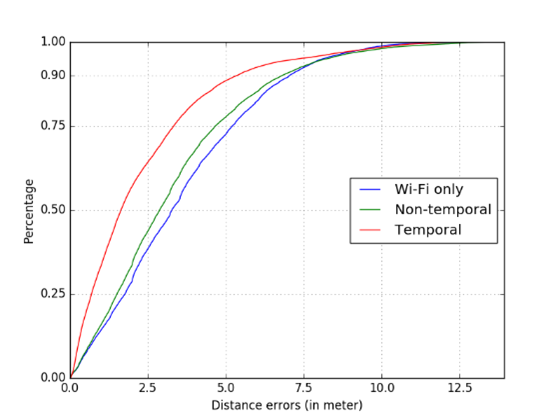}
\caption{Cumulative distance errors for three approaches \label{fig::error1}}
\end{figure}

Figure \ref{fig::error1} illustrates the distance error for three approaches over all the scenarios. Both the \textit{Wi-Fi Only} and the \textit{Non-temporal} have a closed performance. For 75\% of times, the distance errors of two approaches are around 5m. The Bluetooth-based relative distance are employed more efficiently in to \textit{Temporal} approach. It has a significant improvement from the Wi-Fi-based tracking. For 75\% of times and 90\% of times, the errors of \textit{Temporal} approach stay around 3.0m and 5.0m, respectively. Beside the Bluetooth information, adding of map-based information and moving model constraint also reduce noisy output from the standard RF Wi-Fi fingerprinting model.

\begin{figure}[ht!]
	\centering
	\begin{tabular}{cc}
    \subfloat[Device 1]{\includegraphics[width=0.45\linewidth]{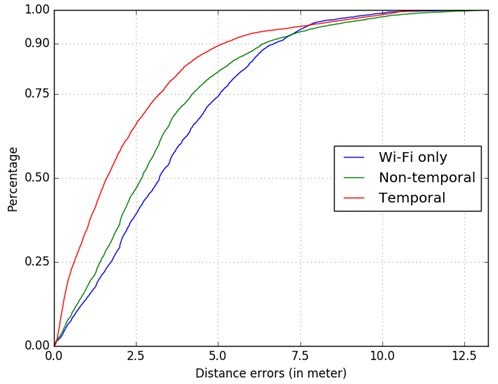}}	& 
	\subfloat[Device 2]{\includegraphics[width=0.45\linewidth]{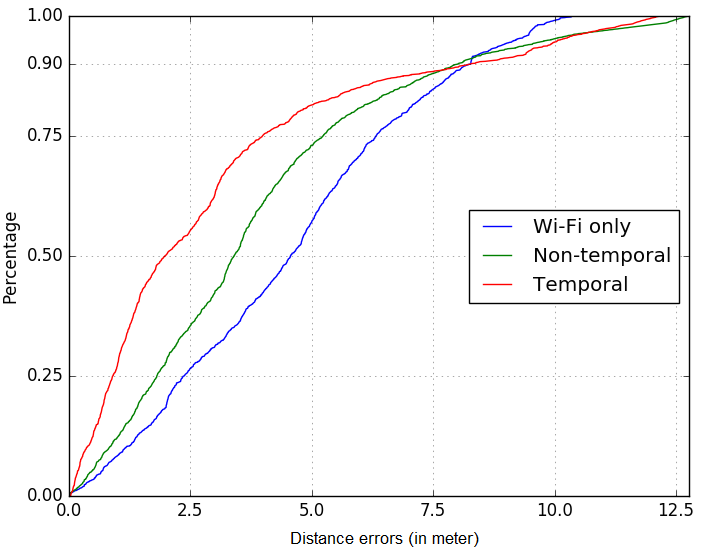}} \\
	\subfloat[Device 3]{\includegraphics[width=0.45\linewidth]{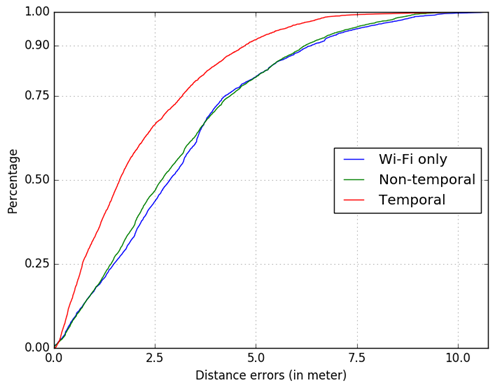}}	& 
    \subfloat[Device 4]{\includegraphics[width=0.45\linewidth]{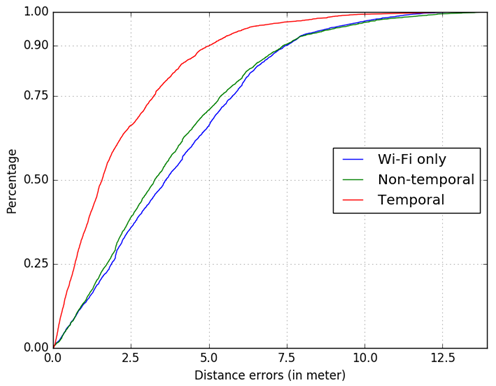}}
	\end{tabular}

     \caption{Cumulative distance errors for each testing devices}\label{fig::error2}
\end{figure}

Individual distribution error for each tested device is given in Figure \ref{fig::error2}. Both smartphones, Samsung Galaxy Note 4 and HTC One ME, have a similar distribution. The \textit{Non-temporal} approach presents a slightly improvement from using only Wi-Fi data and the temporal approach can reduce the error significantly for the regions less than 7.5m. However, the addition of Bluetooth data does not help when the tracking errors exceed 7.5m. The errors of all three models are distributed similarly at the error regions larger than 10m. It even adds more noise to the tracking results of Device 2. In the case of Device 3 and Device 4, both \textit{Non-temporal} and \textit{Wi-Fi Only} have nearly identical distributions and the temporal one outperforms the two others. The \textit{Temporal} has the {highest} improvement with Device 4, which can overcome the issue of non-training data.

\section{Conclusion}
In this work, we have presented a collaborative tracking framework based on the smartphone’s Wi-Fi and Bluetooth scanning data. The Wi-Fi data is used as a raw positioning output, which is then improved by the relative distance from Bluetooth inquiry RSS signals. Two combination approaches are introduced, which is the \textit{Non-temporal} approach and \textit{Temporal} approach. The \textit{Non-temporal} approach attempts to simplify the information fusion task by removing the time-relationship between different Wi-Fi scan and Bluetooth scan. The \textit{Temporal} approach takes a more direct way to establish the conditions between the two types of data. {Both} approaches have been tested and compared with the standard Wi-Fi fingerprinting model. From the testing results, while the \textit{Non-temporal} is only applicable in some specific scenarios, the \textit{Temporal} approach outperforms the Wi-Fi fingerprinting models significantly. {This} study has shown that the collaborative positioning based on the Wi-Fi and Bluetooth data would be applicable in a multi-user context. Combining two type of wireless data can reduce the noise from Wi-Fi fingerprinting model significant. However, the testing scenario are still {dealing with} simple contexts of multiple users. There are also some remaining issues on the technical aspects, such as the communication between the users and the server, energy impact on the smartphone and signal inference between multiple devices.

\section*{Acknowledgement}
This publication is part of the output of the ASEAN IVO project \textit{IoT System for Public Health and Safety Monitoring with Ubiquitous Location Tracking}.

\bibliography{ref}
\bibliographystyle{IEEEtran}

\end{document}